\documentstyle[preprint,aps]{revtex}
\begin{document}
\tightenlines
\title{
\begin{flushright}\normalsize HU-EP-01/27\\ \smallskip RCNP-Th01018 \\
\vskip 1cm
\end{flushright}
Form factors of heavy-to-light  $\bbox{B}$ decays at large
recoil}
\author{D. Ebert$^{1,2}$,
 R. N. Faustov$^{2,3}$ and V. O. Galkin$^{2,3}$ }
\address{$^1$Research Center for Nuclear Physics (RCNP), Osaka University,
Ibaraki, Osaka 567, Japan\\
$^2$Institut f\"ur Physik, Humboldt--Universit\"at
zu Berlin, 10115 Berlin, Germany\\
$^3$Russian Academy of Sciences, Scientific Council for
Cybernetics, Moscow 117333, Russia}

\maketitle

\begin{abstract}
 General 
  relations between the form factors of $B$ decays to light mesons are
  derived using the heavy quark and large recoil expansion. On their
  basis the
  complete account of contributions of second order in the ratio of
  the light meson mass to the large recoil energy is performed. Both
  ground and excited final meson states are considered. It is shown
  that most of the known form factor relations remain valid after the
  inclusion of quadratic mass corrections. The validity of some of
  such relations requires additional equalities for the helicity
  amplitudes. It is found that all these relations and equalities are
  fulfilled in the relativistic quark model based on the
  quasipotential approach in quantum field theory. The contribution of
  $1/m_b$ corrections to the branching fraction of the rare radiative $B$
  decay is discussed.

\medskip
\noindent PACS number(s): 13.20.He, 12.39.Hg, 12.39.Ki
\end{abstract}

\section{Introduction}
\label{sec:int}

The investigation of exclusive weak $B$ decays to  mesons 
composed of the light $u,d,s$  quarks (heavy-to-light decays)
represents  an important problem in particle physics. It provides
means to measure such fundamental parameters as the Cabibbo-Kobayashi-Maskawa
(CKM) matrix elements $V_{ub}$, $V_{td}$ and $V_{ts}$, to predict
CP-violating asymmetries, to test the Standard Model both at large and
short distances and to look for possible manifestations of new physics.
In contrast with  $B$ decays to $D$ mesons, where both
initial and final mesons contain the heavy $b$ and $c$ quarks
(heavy-to-heavy decays) and thus well developed methods of heavy quark
effective theory (HQET) \cite{mw} can be applied,
the heavy-to-light decays are less studied theoretically. HQET and
heavy quark spin-symmetry do not reduce the number of independent form
factors for each decay mode in the latter case, but yield only
relations between form factors of different heavy-to-light modes,
e.~g. semileptonic and rare radiative $B$ decays as in the Isgur-Wise
relations \cite{iw}.
 
In heavy-to-light decays the final meson usually carries a large
recoil momentum (energy) of order of half a $B$ meson mass. The
inverse of this large quantity can serve as an expansion parameter of
corresponding large energy effective theory (LEET). It was used
originally by Dugan and Grinstein \cite{dg} in studying nonleptonic
two-body $B$ decays to a charmed and a light meson (so-called
energetic decays). Further development was given in Ref.~\cite{a}. 

We used a similar approach within the relativistic quark model while
considering exclusive rare radiative $B$ decays to $K^*$ and
$\gamma$ \cite{fg}. For the real photon ($q^2=0$) the $K^*$ recoil momentum
$\Delta$ has a fixed value $(M^2-m_{K^*}^2)/(2M)$ and its
energy is  equal to $(M^2+m_{K^*}^2)/(2M)$. Since the $B$ meson mass 
$M\gg m_{K^*}$ and
$M$ is of order of the $b$ quark mass $m_b$, both the recoil momentum
and energy are of order of $m_b/2$. Thus we can make an expansion in
$1/\Delta\sim 2/m_b$ for the final state and the usual $1/m_b$ heavy
quark expansion for the initial one. These expansions significantly
simplify calculations of the matrix element.  Later we applied  the
same approach to the description of exclusive semileptonic
heavy-to-light $B$ decays \cite{fgm}. In this case the value of the
light meson recoil momentum is not fixed, but still is of order of
$m_b/2$ almost in the hole kinematical range.
The Isgur-Wise relation originally derived for the heavy-to-light form
factors at maximum $q_{\rm max}^2=(M-m_V)^2$ is satisfied in our model for
small $q^2\ll M^2, m_V^2$ (near the point of maximum recoil, $m_V$
is the mass of a final vector meson). 
A close approach was used by Stech \cite{stech} and Soares
\cite{soares} for describing heavy-to-light transitions.     

Recently it has been shown \cite{clopr} that in the heavy quark and
large recoil limit new symmetries emerge and impose new relations on
the form factors (thus reducing the number of independent ones) of
heavy-to-light $B$ decays. The interaction with collinear gluons
preserves these relations, establishing them in the large energy limit
of QCD \cite{bfps}. On the other hand these new symmetries are broken
by radiative corrections \cite{bf}.

In this paper we extend the analysis of Ref.~\cite{clopr} to include
systematically the contributions quadratic in the final to initial
meson mass ratio and of order $q^2/M^2$. Their inclusion is
especially important when the final meson is in the radially or
orbitally excited state and the above contributions become quite
appreciable. To achieve this goal the general expressions for the
heavy-to-light form factors accounting for nonzero final meson mass
are derived. They allow to obtain many important relations between
form factors. In particular, the inclusion of these contributions
is necessary for the exact fulfilment of the Isgur-Wise relations \cite{iw}
and some of the relations given by Soares \cite{soares}. Additional
constraints which are specific for the constituent quark model lead to the
fulfilment of the rest of the relations found by Soares \cite{soares2}.

The paper is organized as follows. In Sec.~\ref{sec:efc} we consider
the effective theory for the description of heavy-to-light decays at large
recoil of the final meson. 
Special emphasis is made on the inclusion
of the corrections of second order in the ratio of the final meson mass
to its recoil energy. Such corrections are especially important for
$B$ decays to excited light mesons since their mass is not small
enough. The general formula for calculating weak decay matrix elements
in the effective theory is given. In Secs.~\ref{sec:rexc} and
\ref{sec:oexc} the effective theory is applied for deriving various symmetry
relations between the form factors of $B$ decays to ground  and
radially or orbitally excited meson states.  A significant
reduction in the number of independent decay form factors in the heavy
quark and large recoil limit is observed. The obtained form factor
relations  are compared to known ones. The
fulfilment of form factor relations in the relativistic quark model is
tested in Sec.~\ref{sec:srrm}. Then, some phenomenological applications
are discussed. Finally, we give our conclusions in
Sec.~\ref{sec:conc}.

\section{The large recoil effective theory}
\label{sec:efc}
We consider the kinematical region where the final meson energy
$E_F=(M^2+m^2-q^2)/(2M)$ is of order $M/2$, $m^2\ll M^2$ and the
four-momentum transfer squared $q^2\ll M^2$. This means that the final
light meson bears a large recoil three-momentum $\Delta\sim M/2$. We 
assume that the main part of this momentum is carried by an active
light quark in the final meson and that its interaction with the
spectator light quark is soft, hence we neglect all hard gluon
contributions. We also neglect the Sudakov suppression of the soft part
of the form factors \cite{clopr,bf}. 
Since the $B$ meson mass $M\sim m_b$, we use the $1/M$
expansion retaining all terms of order $m^2/M^2$ and neglecting the
ones of order $\Lambda_{\rm QCD}/M$ and higher.\footnote{The $m^2/M^2$ 
corrections are especially important for decays to excited light
mesons, since their mass $m$ is in the range $1.2-1.5$ GeV, i.~e. 
about the charmed quark mass.}  
 To this end we
introduce the following kinematical notations:
the four-momentum of the heavy $B$ meson with mass $M$ and velocity
$v$  
\begin{equation}
  p_B=Mv,
\end{equation}
the heavy quark momentum 
\begin{equation}
  p_Q=m_Q v +k,
\end{equation}
where $m_Q$ is the heavy quark mass and $k$ is a small residual momentum
($|k|\sim \Lambda_{\rm QCD}\ll m_Q$).

The energetic final light meson carries a momentum
\begin{equation}
\label{eq:pf}
 p_F=E n +\frac{m^2}{4E}\eta, 
\end{equation}
where we have introduced two light-like vectors $n_\mu$ and
$\eta_\mu=2v_\mu -n_\mu$ ($n^2=0$ and $\eta^2=0$) satisfying the
relations $n\cdot v=1$ and $n\cdot\eta=2$. In the rest frame of the
initial $B$ meson, where $v=(1,0,0,0)$,  choosing the momentum of the
final light meson in $z$ direction,
 we get  $n=(1,0,0,1)$ and
$\eta=(1,0,0,-1)$. It is easy to check that $p_F^2=m^2$, where $m$ is
the mass of the final meson. The important difference with the
case when the final meson
mass is being neglected consists in that $E$ is not the on-shell
energy of the final meson. Indeed, the on-shell energy $E_F$ and the
recoil momentum ${\bf \Delta}$, which form the final meson
four-momentum $p_F=(E_F,{\bf \Delta})$, are related to $E$ by
\begin{eqnarray}
  \label{eq:ef}
  E_F&=&\frac{M^2+m^2-q^2}{2M}=E\left(1+\frac{m^2}{4E^2}\right), \\ 
\label{eq:delt}  
|{\bf \Delta}|&\equiv&\Delta=\sqrt{E_F^2-m^2}=
E\left(1-\frac{m^2}{4E^2}\right),\\
\nonumber 2E&=&E_F+\Delta,\qquad \frac{m^2}{2E}=E_F-\Delta, \qquad q=p_B-p_F.
\end{eqnarray}
Under our assumptions it follows from these  formulas that 
$E_F\sim\Delta\sim E\sim M/2$.
The active light quark momentum in the 
final meson can be represented as
\begin{equation}
 p_q=E n +\frac{m^2}{4E}\eta +k'\equiv\Delta n+\frac{m^2}{2E}v+k', 
\end{equation}
where the
residual momentum $k'$ is small, if we neglect hard gluon
exchanges,  and satisfies $|k'|\sim \Lambda_{\rm QCD}\ll E$ and $|k'|\ll m$.   

The two component light quark fields $q_\pm(x)$ can be defined from
the full QCD fields $q(x)$ by
\begin{equation}
  \label{eq:fq}
  q_\pm(x)=e^{i(\Delta n\cdot x+\frac{m^2}{2E}v\cdot x)}P_\pm q(x),
\end{equation}
where $P_\pm$ are the projectors
\begin{equation}
  \label{eq:proj}
  P_+=\frac{\not\! n \not\! \eta}4=\frac{\not\! n \not\!
  v}2,\qquad   P_-=\frac{\not\! \eta\not\! n}4= \frac{\not\!
  v\not\! n}2.  
\end{equation}
Substituting expressions (\ref{eq:fq}) in the QCD Lagrangian ${\cal
L}_{\rm QCD}=\bar q(i\not\!\! D-m_q)q$, one obtains 
\begin{eqnarray}
  \label{eq:lagr}
  {\cal L}_{\rm QCD}&=&\bar q_+(x) \not\! v\left(i n\cdot D
  +\frac{m^2}{2E}\right) q_+(x) +\bar q_-(x)i \not\!\! D q_+(x)+ \bar
  q_+(x)i \not\!\! D q_-(x)\cr\cr 
&& +\bar q_-(x)\not\! v(2E+2iv\cdot D-in\cdot
  D)q_-(x),
\end{eqnarray}
where the covariant derivative $D^\mu=\partial^\mu -ig_sA^\mu$.
The variation of this Lagrangian over $\delta q_-$ gives the equation of
motion
\begin{equation}
  \label{eq:eqm}
  \frac{\delta{\cal L}_{\rm QCD}}{\delta q_-}=i \not\!\! D q_+(x)+ \not\!
  v(2E+2iv\cdot D-in\cdot D)q_-(x) =0, 
\end{equation}
which allows to express formally $q_-(x)$ in terms of $q_+(x)$
\begin{equation}
  \label{eq:qmin}
  q_-(x)=-\frac{i\not\! v\not\!\! D}{2E+2iv\cdot D-in\cdot D}\ q_+(x).
\end{equation}
Thus $q_-(x)$ is suppressed by $\Lambda_{\rm QCD}/E$ with
respect to $q_+(x)$. Expanding the Lagrangian  ${\cal L}_{\rm QCD}$
(\ref{eq:lagr})  in
inverse powers of $1/E$ up to leading order in $\Lambda_{\rm QCD}/E$
and keeping terms of order $m^2/E^2$, we get the effective Lagrangian
\begin{equation}
  \label{eq:effl}
  {\cal L}_{\rm eff}=\bar q_n(x) \not\! v\left(i n\cdot D
  +\frac{m^2}{2E}\right) q_n(x),
\end{equation}
where $q_n(x)\equiv q_+(x)$. This effective Lagrangian differs from
the ${\cal L}_{\rm LEET}$ one of Ref.~\cite{clopr} by the second
term, which represents the
correction accounting for the final meson mass.
However, this correction does not violate the symmetry of the
leading order Lagrangian, since it has the Dirac structure similar to
the leading contribution. Thus the symmetry relations are not spoiled
by this term.\footnote{The arising invariant functions in 
this case depend  not
only on the recoil energy but on the final meson mass as
well. However,
this dependence does not break the constraints on the form factors
imposed by LEET, since the decay matrix elements for each  final meson
state are described by their
own set of invariant functions which is the same as in LEET.}  
For the complete determination of the $m^2/E^2$
corrections  it is necessary to use the exact expression (\ref{eq:pf})
for the final meson momentum and relations (\ref{eq:ef}), (\ref{eq:delt}) 
between $E$, on-shell energy $E_F$ and recoil momentum $\Delta$ of the
final meson. Taking this into account we obtain for the sum and difference of
four-momenta of the initial and final meson   
\begin{eqnarray}
  \label{eq:ppf}
  p_B+p_F&=&M\left(1+\frac{m^2}{2ME}\right)v +\Delta n,\cr
q\equiv p_B-p_F&=&M\left(1-\frac{m^2}{2ME}\right)v -\Delta n.
\end{eqnarray}
The transversality conditions for the polarization vector
$\epsilon_\mu$  of the (axial) vector mesons and tensor
$\epsilon_{\alpha\beta}$ of the 
tensor mesons $\epsilon^*\cdot p_F=0$  and $\epsilon^*_{\alpha\beta} 
p_F^\beta=0$ imply that
\begin{eqnarray}
  \label{eq:pol}
  \epsilon^*\cdot n&=&-\frac{m^2}{2E\Delta}\epsilon^*\cdot v,\cr\cr
\epsilon^*_{\alpha\beta}n^\beta&=&-\frac{m^2}{2E\Delta}
\epsilon^*_{\alpha\beta}v^\beta.
\end{eqnarray}

The symmetry relations between soft form factors of weak $B$ decays to
excited light mesons in the large recoil limit can be obtained
using methods of Ref.~\cite{bf}. The matrix element of the quark current
between the initial $B$ meson state in the infinitely heavy quark
limit and the final state of the excited light meson at large recoil
is given by
\begin{equation}
  \label{eq:mxelt}
  \langle F(p_F)|\bar q_n \Gamma h_v|B(p)\rangle ={\rm tr}[A_F(E_F)
  \bar{\cal M}_F\Gamma {\cal M}_B],
\end{equation}
where $F=P,V,S,A,T$ for pseudoscalar, vector, scalar, axial vector and
tensor, respectively. The matrices ${\cal M}_F$ and ${\cal M}_B$ have
the form
\begin{equation}
  \label{eq:mfb}
 \bar{\cal M}_F=\left\{
    \begin{array}{c}
-\gamma_5\cr
\not\! \epsilon^*\cr
1\cr
-\not\! \epsilon^*\gamma_5\cr
\epsilon^*_{\mu\nu}\gamma^\mu v^\nu 
\end{array}
\right\}\frac{\not\! v \not\! n}2 \hskip 1.5cm
\begin{array}{c}
F=P \cr F=V\cr F=S\cr F=A\cr F=T\cr
\end{array}
,\hskip 1.5cm
{\cal M}_B=-\frac{1+\not\! v}2\gamma_5,
\end{equation}
where $\epsilon_\mu$ and  $\epsilon_{\mu\nu}$ are the polarization vector
and tensor of the (axial) vector and tensor mesons, respectively. 
The functions
$A_F(E_F)$ are independent of the Dirac structure $\Gamma$ of
the current, since there are no dynamical (i.~e. contracted with the
covariant derivative $D$)  Dirac matrices in the effective
Lagrangian (\ref{eq:effl}), and parametrize the long-distance
dynamics. The most general form of $A_F(E_F)$ is \cite{bf}
\begin{equation}
  \label{eq:alf}
  A_F(E_F)=a_{1F}(E_F)+a_{2F}(E_F)\not\! v+a_{3F}(E_F)\not\! n+
  a_{4F}(E_F)\not\! n \not\! v.
\end{equation}
However, the presence of projection operators imply that not all the
functions  $a_{iF}(E_F)$ are independent. As the result,
 the following
parametrization of $A_F(E_F)$ can be obtained
\begin{eqnarray}
  \label{eq:axip}
  A_{P,S}(E_F)&=&2E_F \zeta_{P,S}(E_F)\\
\label{eq:axiv}
A_{V,A,T}(E_F)&=&E_F\not\!
n\left[\zeta_{V,A,T}^\perp(E_F)-\frac{\not\! v}2\frac{m}{E_F}
  \zeta_{V,A,T}^\parallel(E_F)\right].  
\end{eqnarray}
Our definition of functions  $\zeta_{P}(E_F)$,
$\zeta_{V}^\perp(E_F)$ and $\zeta_{V}^\parallel(E_F)$ 
coincides
with that of Ref.~\cite{clopr}, but differs from \cite{bf}  by an
extra factor $m/E_F$  in front of
$\xi_{V,A,T}^\parallel(E_F)$.

\section{$\bbox{B}$ decays to ground state and 
radially excited  light  mesons}
\label{sec:rexc}
First we consider the heavy-to-light $B$ decays to ground state and
radially  excited  pseudoscalar and vector light mesons.
 
\subsection{$\bbox{B}$ decays to  pseudoscalar light  mesons}
\label{sec:ps}

The matrix elements of $B$ decays to pseudoscalar  mesons
can be parametrized by three invariant form factors:
\begin{eqnarray}
  \label{eq:pff1}
  \langle P(p_F)|\bar q \gamma^\mu b|B(p_B)\rangle
  &=&f_+(q^2)\left[p_B^\mu+ p_F^\mu-\frac{M^2-m_P^2}{q^2}\ q^\mu\right]+
  f_0(q^2)\frac{M^2-m_P^2}{q^2}\ q^\mu,\\ \cr
\label{eq:pff2}
\langle P(p_F)|\bar q \sigma^{\mu\nu}q_\nu b|B(p_B)\rangle&=&
\frac{if_T(q^2)}{M+m_P} [q^2(p_B^\mu+p_F^\mu)-(M^2-m_P^2)q^\mu],
\end{eqnarray}
where $f_+(0)=f_0(0)$; $M$ is the $B$ meson mass 
and $m_P$ is the pseudoscalar meson mass.

We obtain the symmetry relations for the corresponding form factors by
evaluating the trace in (\ref{eq:mxelt}) with the account of
(\ref{eq:axip}), (\ref{eq:ppf}), (\ref{eq:pol}) and find the
following relations
\begin{eqnarray}
  \label{eq:srp1}
  \langle P(p_F)|\bar q \gamma^\mu b|B(p_B)\rangle&=& 2E_F
  \zeta_{P}(E_F)n^\mu,\\ \cr
 \label{eq:srp2}
\langle P(p_F)|\bar q \sigma^{\mu\nu}q_\nu b|B(p_B)\rangle&=&2iE_F
\zeta_{P}(E_F)
\left[(M-E_F)n^\mu-M\left(1-\frac{m_P^2}{2EM}\right)v^\mu\right]. 
\end{eqnarray}
The factor in front of $v^\mu$ in (\ref{eq:srp2}) can be expressed in terms
of observable variables $E_F$ and $\Delta$ with the help of Eqs.~(\ref{eq:ef}) 
and (\ref{eq:delt})
\begin{equation}
  \label{eq:me}
  \frac{m_P^2}{2EM}=\frac{E_F-\Delta}M.
\end{equation}
Thus all form factors for the $B$ decay to a pseudoscalar meson in the heavy
quark and large recoil limit can be parametrized by one 
function $ \zeta_{P}(E_F)$. Comparing Eqs.~(\ref{eq:pff1})--(\ref{eq:srp2})
we find the following expressions for the form factors in terms of this 
invariant function
\begin{eqnarray}
  \label{eq:ffpe}
  f_+(q^2)&=&\left(1-\frac{m_P^2}{2EM}\right)\frac{E_F}{\Delta}
\zeta_{P}(E_F),\cr\cr
f_0(q^2)&=&\frac{2E_FM}{M^2-m_P^2}\left(1-\frac{m_P^2}{2EM}\right)
\zeta_{P}(E_F), \cr\cr
 f_T(q^2)&=&\frac{M+m_P}{M}\frac{E_F}{\Delta}
 \zeta_{P}(E_F).
\end{eqnarray}
Taking into account that our consideration is valid up to second order in 
$m_P/E$ and all other
corrections were neglected, we omit terms of fourth order 
in $m_P/E_F$ and terms of order $q^2 m_P^2/E_F^4$ in
Eq.~(\ref{eq:ffpe})  and finally get
\begin{eqnarray}
\Delta\cong \frac{M^2-m_p^2-q^2}{2M},&&\quad \frac{m_P^2}{2EM}\cong
\frac{m_P^2}{M^2}, \qquad \frac{E_F}\Delta\cong 1+\frac{2m_P^2}{M^2},\\\cr
  \label{eq:ffp}
  f_+(q^2)&=&\left(1+\frac{m_P^2}{M^2}\right)\zeta_{P}(E_F),\cr\cr
f_0(q^2)&=&\frac{2E_F}M
\zeta_{P}(E_F), \cr\cr
 f_T(q^2)&=&\frac{M+m_P}{M}\left(1+\frac{2m_P^2}{M^2}\right)
 \zeta_{P}(E_F). 
\end{eqnarray}
As a result,
 the following relations between the form factors of
$B$ decays to scalar light mesons in the heavy quark and large recoil 
limit  are obtained
\begin{eqnarray}
\label{eq:ffrp}
 f_+(q^2)&=&\frac{M}{2E_F}\left(1+\frac{m_P^2}{M^2}\right)f_0(q^2)=
 \frac{M}{M+m_P}\left(1-\frac{m_P^2}{M^2}\right)f_T(q^2)\cr\cr
&&=
\left(1+\frac{m_P^2}{M^2}\right)\zeta_{P}(E_F).
\end{eqnarray}

\subsection{$\bbox{B}$ decays to vector  light  mesons}
\label{sec:vc}

The matrix elements of weak current for $B$ decays to vector mesons
are parametrized by seven form factors
\begin{eqnarray}
  \label{eq:vff1}
  \langle V(p_F)|\bar q \gamma^\mu b|B(p_B)\rangle&=
  &\frac{2iV(q^2)}{M+m_V} \epsilon^{\mu\nu\rho\sigma}\epsilon^*_\nu
  p_{B\rho} p_{F\sigma},\\ \cr
\label{eq:vff2}
\langle V(p_F)|\bar q \gamma^\mu\gamma_5 b|B(p_B)\rangle&=&2m_V
A_0(q^2)\frac{\epsilon^*\cdot q}{q^2}\ q^\mu
 +(M+m_V)A_1(q^2)\left(\epsilon^{*\mu}-\frac{\epsilon^*\cdot
    q}{q^2}\ q^\mu\right)\cr\cr
&&-A_2(q^2)\frac{\epsilon^*\cdot q}{M+m_V}\left[p_B^\mu+
  p_F^\mu-\frac{M^2-m_V^2}{q^2}\ q^\mu\right], \\\cr
\label{eq:vff3}
\langle V(p_F)|\bar q i\sigma^{\mu\nu}q_\nu b|B(p_B)\rangle&=&2T_1(q^2)
\epsilon^{\mu\nu\rho\sigma} \epsilon^*_\nu p_{F\rho} p_{B\sigma},\\\cr
\label{eq:vff4}
\langle V(p_F)|\bar q i\sigma^{\mu\nu}\gamma_5q_\nu b|B(p_B)\rangle&=&
T_2(q^2)[(M^2-m_V^2)\epsilon^{*\mu}-(\epsilon^*\cdot q)(p_B^\mu+
p_F^\mu)]\cr\cr
&&+T_3(q^2)(\epsilon^*\cdot q)\left[q^\mu-\frac{q^2}{M^2-m_V^2}
  (p_B^\mu+p_F^\mu)\right], 
\end{eqnarray}
where $2m_VA_0(0)=(M+m_V)A_1(0)-(M-m_V)A_2(0)$, $T_1(0)=T_2(0)$; 
$m_V$ and $\epsilon_\mu$ are the mass and polarization vector of
the final vector meson. 
Calculating corresponding traces in (\ref{eq:mxelt}) and accounting for
(\ref{eq:axiv}), (\ref{eq:ppf}), (\ref{eq:pol}), we get the
heavy quark and large recoil symmetry relations 
\begin{eqnarray}
  \label{eq:srv1}
 \langle V(p_F)|\bar q \gamma^\mu b|B(p_B)\rangle&=&2iE_F
 \zeta_V^\perp(E_F)\epsilon^{\mu\nu\rho\sigma}\epsilon^*_\nu
  v_\rho n_\sigma,\\\cr
 \label{eq:srv2}
\langle V(p_F)|\bar q \gamma^\mu\gamma_5 b|B(p_B)\rangle&=&2E_F\Biggl\{
  \zeta_V^\perp(E_F)
\left[\epsilon^{*\mu}-\epsilon^*\cdot v \left(\frac{E_F}{\Delta}n^\mu
-\frac{m_V^2}{2E\Delta}v^\mu\right)\right]\cr\cr
&&
+\frac{E}{\Delta} \frac{m_V}{E_F}\zeta_V^\parallel(E_F)\epsilon^*\cdot v
  n^\mu\Biggr\},\\\cr
  \label{eq:srv3}
\langle V(p_F)|\bar q i\sigma^{\mu\nu}q_\nu b|B(p_B)\rangle&=&2iE_F M
\zeta_V^\perp(E_F) \left(1-\frac{m_V^2}{2EM}\right)
\epsilon^{\mu\nu\rho\sigma} \epsilon^*_\nu n_\rho v_\sigma,\\\cr
\label{eq:srv4}
\langle V(p_F)|\bar q i\sigma^{\mu\nu}\gamma_5q_\nu
b|B(p_B)\rangle&=&2E_F\Biggl\{M\zeta_V^\perp(E_F)
  \left(1-\frac{m_V^2}{2EM}\right) \Biggl[\epsilon^{*\mu}
  -\epsilon^*\cdot v \Biggl(\frac{E_F}{\Delta}n^\mu\cr\cr
&&-\frac{m_V^2}{2E\Delta}v^\mu\Biggr)\Biggr]
 +\frac{E}{\Delta}\frac{m_V}{E_F}\zeta_V^\parallel(E_F)
  \epsilon^*\cdot v \Biggl[(M-E_F)n^\mu \cr\cr
&& -M
    \left(1-\frac{m_V^2}{2EM}\right)v^\mu\Biggr]\Biggr\},  
\end{eqnarray}
All form factors for $B$ decays to vector mesons in the heavy
quark and large recoil limit can be expressed through two
invariant functions $\zeta_V^\perp(E_F)$ and $\zeta_V^\parallel(E_F)$.
Comparing the invariant decompositions 
(\ref{eq:vff1})--(\ref{eq:vff4})
with the
symmetry relations (\ref{eq:srv1})--(\ref{eq:srv4}), we get the
following expressions  
\begin{eqnarray}
  \label{eq:ffve}
  V(q^2)&=&\frac{M+m_V}M\frac{E_F}{\Delta}\zeta_V^\perp(E_F),\cr\cr
A_1(q^2)&=&\frac{2E_F}{M+m_V}\zeta_V^\perp(E_F),\cr\cr
A_2(q^2)&=&\frac{M+m_V}M\frac{E_F^2}{\Delta^2}
\left[\left(1-\frac{m_V^2}{E_FM}\right)
\zeta_V^\perp(E_F)
-\left(1-\frac{m_V^2}{2EM}\right)
\frac{Em_V}{E_F^2}\zeta_V^\parallel(E_F)\right],\cr\cr
A_0(q^2)&=&\left(1-\frac{m_V^2}{2EM}\right)
\frac{E}{\Delta}\zeta_V^\parallel(E_F),\cr\cr
T_1(q^2)&=&\left(1-\frac{m_V^2}{2EM}\right)
\frac{E_F}{\Delta}\zeta_V^\perp(E_F),\cr\cr
T_2(q^2)&=&\frac{2E_FM}{M^2-m_V^2}\left(1-\frac{m_V^2}{2EM}\right)
 \zeta_V^\perp(E_F),\cr\cr
T_3(q^2)&=&\frac{E_F^2}{\Delta^2}\left[\left(1+\frac{m_V^2}{E_FM}\right)
\left(1-\frac{m_V^2}{2EM}\right)\zeta_V^\perp(E_F)-
\left(\frac{M^2-m_V^2}{M^2}\right)
\frac{Em_V}{E_F^2}\zeta_V^\parallel(E_F)\right].  
\end{eqnarray}

Within the accuracy of our consideration we further expand 
expressions (\ref{eq:ffve}) up to terms of order $m_V^2/M^2$  
\begin{eqnarray}
  \label{eq:ffv}
  V(q^2)&=&\frac{M+m_V}M\frac{E_F}{\Delta}\zeta_V^\perp(E_F),\cr\cr
A_1(q^2)&=&\frac{2E_F}{M+m_V}\zeta_V^\perp(E_F),\cr\cr
A_2(q^2)&=&\frac{M+m_V}M\left(1+\frac{2m_V^2}{M^2}\right)\left[
\zeta_V^\perp(E_F)-\frac{m_V}{E_F}\zeta_V^\parallel(E_F)\right],\cr\cr
A_0(q^2)&=&\zeta_V^\parallel(E_F),\cr\cr
T_1(q^2)&=&\left(1-\frac{m_V^2}{M^2}\right)
\frac{E_F}{\Delta}\zeta_V^\perp(E_F),\cr\cr
T_2(q^2)&=&\frac{2E_F}M \zeta_V^\perp(E_F),\cr\cr
T_3(q^2)&=&\left(1+\frac{5m_V^2}{M^2}\right)\zeta_V^\perp(E_F)-
\left(1+\frac{2m_V^2}{M^2}\right)\frac{m_V}{E_F}\zeta_V^\parallel(E_F).  
\end{eqnarray}
From Eqs.~(\ref{eq:ffv}) it is possible to deduce the following
relations between the heavy-to-light form factor for the case of the
final vector meson 
\begin{eqnarray}
  \label{eq:ffrv1}
  \frac{M}{M+m_V}\frac{\Delta}{E_F}V(q^2)&=&
\frac{M+m_V}{2E_F}A_1(q^2)=
\frac{M}{M+m_V}\left(1-\frac{2m_V^2}{M^2}\right)A_2(q^2)
+\frac{m_V}{E_F}A_0(q^2)\cr\cr
&=&\left(1+\frac{m_V^2}{M^2}\right)\frac{\Delta}{E_F}
T_1(q^2)=\frac{M}{2E_F}T_2(q^2)
=\zeta_V^\perp(E_F),\\\cr
\label{eq:ffrv2}
\frac{m_V}{E_F}A_0(q^2)&=& \frac{M+m_V}{2E_F}A_1(q^2)-
\frac{M}{M+m_V}\left(1-\frac{2m_V^2}{M^2}\right) A_2(q^2)\cr\cr 
&=&\left(1+\frac{2m_V^2}{M^2}\right)T_1(q^2)-
\left(1-\frac{2m_V^2}{M^2}\right)T_3(q^2)\cr\cr 
& =&\frac{M}{2E_F}
\left(1+\frac{3m_V^2}{M^2}\right) T_2(q^2)
-\left(1-\frac{2m_V^2}{M^2}\right)T_3(q^2)=
\frac{m_V}{E_F}\zeta_V^\parallel(E_F). 
\end{eqnarray}
These relations differ by terms of second order in $m_V/E_F$ from the
corresponding relations of Refs.~\cite{clopr,bf} where only some of
$m_V^2/E_F^2$  terms were taken into account. In
particular Eq.~(\ref{eq:ffrv1}) leads to the ratios of form factors
$V(q^2)$ to $A_1(q^2)$ and $T_1(q^2)$ to $T_2(q^2)$, 
which depend only on meson masses and energies
\begin{eqnarray}
  \label{eq:var}
  \frac{V(q^2)}{A_1(q^2)}&=&\frac{(M+m_V)^2}{2M\Delta}=
\frac{(M+m_V)^2}{2M\sqrt{E_F^2-m_V^2}},\cr\cr
 \frac{T_1(q^2)}{T_2(q^2)}&=&\frac{M^2-m_V^2}{2M\Delta}=
\frac{M^2-m_V^2}{2M\sqrt{E_F^2-m_V^2}}.
\end{eqnarray}
Note that the complete account of the $m_V^2/E_F^2$ corrections
results in the replacement of $E_F$ by $\Delta=\sqrt{E_F^2-m_V^2}$ 
in the analogous relation 
(109)  of Ref.~\cite{clopr}. Equations~(\ref{eq:var}) ensure exact vanishing
of the transverse helicity $\lambda=+1$ contribution to the rates of  
$B\to Ve\nu$ and $B\to Vl\bar l$ decays, since the corresponding helicity
amplitudes are given by
\begin{eqnarray}
  \label{eq:helamp}
  H_\pm^{B\to Ve\nu}(q^2)&=&\frac{2M\Delta}{M+m_V}\left[V(q^2)\mp
\frac{(M+m_V)^2}{2M\Delta}A_1(q^2)\right],\cr\cr
H_\pm^{B\to Vl\bar l}(q^2)&=&{2M\Delta}\left[T_1(q^2)\mp
\frac{M^2-m_V^2}{2M\Delta}T_2(q^2)\right].
\end{eqnarray}
Such behaviour of helicity amplitudes is the consequence of the $(V-A)$
structure of weak currents in the standard model, which results in the creation
of a left-handed light quark in the ultra relativistic limit, and the 
helicity-flip amplitude is suppressed by the factor $\Lambda_{\rm
  QCD}/E_F$ \cite{bh}. 
Therefore the helicity $\lambda=-1$ amplitude,  e.~g. for 
$B\to Ve\nu$ decay, can be expressed in the form
\begin{equation}
  \label{eq:hma}
  H_-^{B\to Ve\nu}(q^2)=2(M+m_V)A_1(q^2).
\end{equation}
The helicity $\lambda=0$ amplitude for this decay is given by
\begin{equation}
  \label{eq:h0a}
  H_0^{B\to Ve\nu}(q^2)=\frac1{2m_V\sqrt{q^2}}\left[(M+m_V)
(M^2-m_V^2-q^2)A_1(q^2)-\frac{4M^2\Delta^2}{M+m_V}A_2(q^2)\right]
\end{equation}
and can be rewritten with the account of symmetry relations
(\ref{eq:ffve})  in a more simple form 
\begin{equation}
  \label{eq:hoaa}
  H_0^{B\to Ve\nu}(q^2)=\frac{2M\Delta}{\sqrt{q^2}}A_0(q^2).
\end{equation}
Then the ratio of helicity $\lambda=-1$ and $\lambda=0$  amplitudes reads 
\begin{equation}
  \label{eq:hpmr}
  \frac{|H_-(q^2)|}{|H_0(q^2)|}=\frac{M+m_V}M\frac{\sqrt{q^2}}\Delta
\frac{A_1(q^2)}{A_0(q^2)}=\frac{2\sqrt{q^2}}{M-E_F+\Delta}
\frac{E_F}{E}\frac{|\zeta_V^\perp(E_F)|}{|\zeta_V^\parallel(E_F)|}.
\end{equation}

The Isgur-Wise relations \cite{iw,bd,bh} between the  heavy-to-light
form factors follow from the heavy-quark spin symmetry 
alone, and were obtained near the point of zero recoil 
($\Delta\approx 0$, $q^2\approx q^2_{\rm max}=(M-m_V)^2$) of the final
light meson in the heavy quark limit
\begin{eqnarray}
  \label{eq:iwr}
&&\frac{2M}{M+m_P}f_T(q^2)=f_+(q^2)+\frac{M^2-m_P^2}{q^2}
\left[f_+(q^2)-f_0(q^2)\right],\cr\cr 
&&  T_1(q^2)=\frac{M^2+q^2-m_V^2}{2M}\frac{V(q^2)}{M+m_V}
+\frac{M+m_V}{2M}A_1(q^2),\cr\cr
&&\frac{M^2-m_V^2}{q^2}\left[T_1(q^2)-T_2(q^2)\right]
=\frac{3M^2-q^2+m_V^2}{2M}\frac{V(q^2)}{M+m_V}
-\frac{M+m_V}{2M}A_1(q^2),\cr\cr
&&T_3(q^2)=\frac{M^2-q^2+3m_V^2}{2M}\frac{V(q^2)}{M+m_V}+
\frac{M^2-m_V^2}{Mq^2}m_VA_0(q^2)\cr\cr
&&\qquad\qquad
-\frac{M^2+q^2-m_V^2}{2Mq^2}\left[(M+m_V)A_1(q^2)-(M-m_V)A_2(q^2)\right].
\end{eqnarray}
They are exactly satisfied by form factors (\ref{eq:ffpe}) and 
(\ref{eq:ffve}). Thus we conclude 
that these relations remain valid  near the point $q^2=0$
corresponding to  the maximum recoil  of the
final light meson if the large recoil limit is used in addition. Note
that the Isgur-Wise relations (\ref{eq:iwr}) hold in these limits for 
$B$ decays both to ground state vector (pseudoscalar) light mesons
and their radial excitations.     

Some heavy-to-light form factor relations for decays
to pseudoscalar and vector mesons were obtained by 
Stech \cite{stech} and Soares \cite{soares} in the framework of the 
constituent quark model. 
It is easy to check that the additional relations found at the large 
recoil momentum of the final meson \cite{soares}
\begin{eqnarray}
  \label{eq:soar}
  f_0(q^2)&=&f_+(q^2)\left(1-\frac{q^2}{M^2-m_V^2}\frac{M+E_F-\Delta}
{M-E_F+\Delta}\right),\cr\cr
V(q^2)&=&\frac{(M+m_V)^2}{2M\Delta}A_1(q^2)=\frac{M}{ME_F-m_V^2}
\left[\Delta A_2(q^2)+\frac{M+m_V}Mm_VA_0(q^2)\right]
\end{eqnarray}
are also satisfied exactly by form factors (\ref{eq:ffpe}) and 
(\ref{eq:ffve}). The other relations \cite{soares2} were
derived in the spectator approximation only. Instead of them we get
\begin{eqnarray}
  \label{eq:soares2}
\frac{E_F}E
\frac{\zeta_V^\perp(E_F)}{\zeta_V^\parallel(E_F)}
&=& \frac{T_1(q^2)}{A_0(q^2)}=\frac{m_V(M-m_V)T_2(q^2)}{(ME_F-m_V^2)A_1(q^2)
-\left[2M^2\Delta^2/(M+m_V)^2\right]A_2(q^2)}\cr\cr\cr
&=&\frac{m_V(M+m_V)A_1(q^2)}{(ME_F+m_V^2)T_2(q^2)
-\left[2M^2\Delta^2/(M^2-m_V^2)\right]T_3(q^2)}.
\end{eqnarray}
Thus their fulfilment in the form given in \cite{soares2} requires 
the specific relation between invariant functions 
\begin{eqnarray}
  \label{eq:ffrs1}
 \zeta_V^\parallel(E_F)=\frac{E_F}E\zeta_V^\perp(E_F). 
\end{eqnarray}

\section{$\bbox{B}$ decays to orbitally excited  light  mesons}
\label{sec:oexc}
Now we investigate the heavy-to-light $B$ decays to orbitally excited
$P$-wave light mesons.

\subsection{$\bbox{B}$ decays to  scalar light  mesons}
\label{sec:sc}

The matrix elements of the weak current for $B$ decays to orbitally
excited scalar light mesons can be parametrized by three invariant
form factors
\begin{eqnarray}
  \label{eq:sff1}
  \langle S(p_F)|\bar q \gamma^\mu\gamma_5 b|B(p_B)\rangle
  &=&r_+(q^2)\left(p_B^\mu+ p_F^\mu\right)+
  r_-(q^2)\left(p_B^\mu- p_F^\mu\right),\\\cr
\label{eq:sff2}
\langle S(p_F)|\bar q \sigma^{\mu\nu}\gamma_5 q_\nu b|B(p_B)\rangle&=&
\frac{ir_T(q^2)}{M+m_S} [q^2(p_B^\mu+p_F^\mu)-(M^2-m_P^2)q^\mu],
\end{eqnarray}
where $m_S$ is the scalar meson mass.

The symmetry relations for the matrix elements can be obtained by
calculating corresponding traces in Eq.~(\ref{eq:mxelt}) 
\begin{eqnarray}
  \label{eq:srs1}
  \langle S(p_F)|\bar q \gamma^\mu\gamma_5 b|B(p_B)\rangle&=& 2E_F
  \zeta_{S}(E_F)n^\mu,\\\cr
 \label{eq:srs2}
\langle S(p_F)|\bar q \sigma^{\mu\nu}\gamma_5q_\nu b|B(p_B)\rangle&=&2E_F
\zeta_{S}(E_F)
\left[(M-E_F)n^\mu-M\left(1-\frac{m_S^2}{2EM}v^\mu\right)\right]. 
\end{eqnarray}
These equations show that the form factors can be parametrized by one
invariant function only in the heavy quark and large recoil limit. Comparing
Eqs.~(\ref{eq:sff1}), (\ref{eq:sff2}) and (\ref{eq:srs1}), (\ref{eq:srs2})
we find
\begin{eqnarray}
\label{eq:ffse}
  r_+(q^2)&=&\left(1-\frac{m_S^2}{2EM}\right)\frac{E_F}{\Delta}
\zeta_{S}(E_F),\cr\cr
r_-(q^2)&=&-\left(1+\frac{m_S^2}{2EM}\right)\frac{E_F}{\Delta}
 \zeta_{S}(E_F), \cr\cr
 r_T(q^2)&=&\frac{M+m_S}{M}\frac{E_F}{\Delta}
 \zeta_{S}(E_F). 
\end{eqnarray}
Expanding Eqs.~(\ref{eq:ffse}) in $m_S/E$, we get retaining the terms
of order $m_S^2/M^2$ 
\begin{eqnarray}
  \label{eq:ffs}
  r_+(q^2)&=&\left(1+\frac{m_S^2}{M^2}\right)\zeta_{S}(E_F),\cr\cr
r_-(q^2)&=&-\left(1+\frac{3m_S^2}{M^2}\right) \zeta_{S}(E_F), \cr\cr
 r_T(q^2)&=&\frac{M+m_S}{M}\left(1+\frac{2m_S^2}{M^2}\right)
 \zeta_{S}(E_F). 
\end{eqnarray}
These expressions for the form factors lead to the following 
symmetry relations between them
\begin{equation}
  \label{eq:ffrs}
  r_+(q^2)=-\left(1-\frac{2m_S^2}{M^2}\right)r_-(q^2)=
 \frac{M}{M+m_S}\left(1-\frac{m_S^2}{M^2}\right)r_T(q^2)=
\left(1+\frac{m_S^2}{M^2}\right)\zeta_{S}(E_F). 
 \end{equation}

\subsection{$\bbox{B}$ decays to  axial vector light  mesons}
\label{sec:av}

The matrix elements of the weak current for $B$ decays to axial vector mesons
can be expressed in terms of seven invariant form factors
\begin{eqnarray}
  \label{eq:avff1}
  \langle A(p_F)|\bar q \gamma^\mu b|B(p_B)\rangle&=&
  (M+m_A)t_{V_1}(q^2)\epsilon^{*\mu}
  +[t_{V_2}(q^2)p_B^\mu+t_{V_3}(q^2)p_F^\mu]\frac{\epsilon^*\cdot q}{M} ,\\\cr
\label{eq:avff2}
\langle A(p_F)|\bar q \gamma^\mu\gamma_5 b|B(p_B)\rangle&=&
\frac{2it_A(q^2)}{M+m_A} \epsilon^{\mu\nu\rho\sigma}\epsilon^*_\nu
  p_{B\rho} p_{F\sigma} \\\cr
\label{eq:avff3}
\langle A(p_F)|\bar q i\sigma^{\mu\nu}q_\nu b|B(p_B)\rangle&=&
t_{+}(q^2)\left[(\epsilon^*\cdot q)(p_B+p_F)^\mu -
\epsilon^{*\mu}(M^2-m_A^2)  \right] \cr\cr
&& +t_{-}(q^2)\left[(\epsilon^*\cdot q) q^\mu
-\epsilon^{*\mu} q^2\right] \cr\cr
& &+t_0(q^2)\frac{\epsilon^*\cdot q}{M^2}\left[(M^2-m_A^2)q^\mu
-q^2(p_B+p_F)^\mu \right],\\\cr
\label{eq:avff4}
\langle A(p_F)|\bar q i\sigma^{\mu\nu}\gamma_5q_\nu b|B(p_B)\rangle&=&
2it_+(q^2)\epsilon^{\mu\nu\rho\sigma}\epsilon^*_\nu
  p_{B\rho} p_{F\sigma}, 
\end{eqnarray}
where $m_A$ and $\epsilon^\mu$ are the mass and polarization vector of 
the axial vector meson.

Equation~(\ref{eq:mxelt}) yields the following symmetry equations in this case 
\begin{eqnarray}
  \label{eq:srav1}
 \langle A(p_F)|\bar q \gamma^\mu b|B(p_B)\rangle&=&2E_F\Biggl\{
  \zeta_A^\perp(E_F)\left[\epsilon^{*\mu}-
\epsilon^*\cdot v \left(\frac{E_F}{\Delta}n^\mu
-\frac{m_A^2}{2E\Delta}v^\mu\right)\right]\cr\cr
&&
+\frac{E}{\Delta}
  \frac{m_A}{E_F}\zeta_A^\parallel(E_F)\epsilon^*\cdot v
  n^\mu\Biggr\},\\\cr
 \label{eq:srav2}
\langle A(p_F)|\bar q \gamma^\mu\gamma_5 b|B(p_B)\rangle&=&2iE_F
 \zeta_A^\perp(E_F)\epsilon^{\mu\nu\rho\sigma}\epsilon^*_\nu
  v_\rho n_\sigma,\\\cr
  \label{eq:srav3}
\langle A(p_F)|\bar q i\sigma^{\mu\nu}q_\nu b|B(p_B)\rangle&=&-2E_F
\Biggl\{M\zeta_A^\perp(E_F)
  \left(1-\frac{m_A^2}{2EM}\right) \Biggl[\epsilon^{*\mu}
  -\epsilon^*\cdot v \Biggl(\frac{E_F}{\Delta}n^\mu\cr\cr
&&-\frac{m_A^2}{2E\Delta}v^\mu\Biggr)\Biggr]
 +\frac{E}{\Delta}\frac{m_A}{E_F}\zeta_A^\parallel(E_F)
  \epsilon^*\cdot v \Biggl[(M-E_F)n^\mu \cr\cr
&& -M
    \left(1-\frac{m_A^2}{2EM}\right)v^\mu\Biggr]\Biggr\},\\\cr
\label{eq:srav4}
\langle A(p_F)|\bar q i\sigma^{\mu\nu}\gamma_5q_\nu
b|B(p_B)\rangle&=&2iE_F M
\zeta_A^\perp(E_F) \left(1-\frac{m_A^2}{2EM}\right)
\epsilon^{\mu\nu\rho\sigma} \epsilon^*_\nu v_\rho n_\sigma.  
\end{eqnarray}
Thus all form factors for $B$ decays to axial vector mesons in the heavy
quark and large recoil limit can be expressed through two
invariant functions
\begin{eqnarray}
  \label{eq:ffave}
  t_A(q^2)&=&\frac{M+m_A}M\frac{E_F}{\Delta}\zeta_A^\perp(E_F),\cr\cr
t_{V_1}(q^2)&=&\frac{2E_F}{M+m_A}\zeta_A^\perp(E_F),\cr\cr
t_{V_2}(q^2)&=&\frac{E_Fm_A^2}{\Delta^2 M}\left[2\zeta_A^\perp(E_F)-
\frac{m_A}{E_F}\zeta_A^\parallel(E_F)\right],\cr\cr
t_{V_3}(q^2)&=&2\frac{E_F^2}{\Delta^2}
\left[\frac{Em_A}{E_F^2}\zeta_A^\parallel(E_F)
-\zeta_A^\perp(E_F)\right],\cr\cr
t_+(q^2)&=&\left(1-\frac{m_A^2}{2EM}\right)\frac{E_F}{\Delta}
\zeta_A^\perp(E_F),\cr\cr
t_-(q^2)&=&-\left(1+\frac{m_A^2}{2EM}\right)\frac{E_F}{\Delta}
 \zeta_A^\perp(E_F),\cr\cr
t_0(q^2)&=&\frac{E_FE}{\Delta^2}\left[\frac{m_A}{E_F}
\zeta_A^\parallel(E_F)-\frac{m_A^2}{2E^2}\zeta_A^\perp(E_F)\right].
\end{eqnarray}
Keeping the terms of order $m_A^2/M^2$ in Eqs.~(\ref{eq:ffave}) we
find
\begin{eqnarray}
  \label{eq:ffav}
  t_A(q^2)&=&\frac{M+m_A}M\frac{E_F}{\Delta}\zeta_A^\perp(E_F),\cr\cr
t_{V_1}(q^2)&=&\frac{2E_F}{M+m_A}\zeta_A^\perp(E_F),\cr\cr
t_{V_2}(q^2)&=&\frac{2m_A^2}{M^2}\left[2\zeta_A^\perp(E_F)-
\frac{m_A}{E_F}\zeta_A^\parallel(E_F)\right],\cr\cr
t_{V_3}(q^2)&=&2\left(1+\frac{4m_A^2}{M^2}\right)
\left[\left(1-\frac{m_A^2}{M^2}\right)\frac{m_A}{E_F}\zeta_A^\parallel(E_F)
-\zeta_A^\perp(E_F)\right],\cr\cr
t_+(q^2)&=&\left(1-\frac{m_A^2}{M^2}\right)
\frac{E_F}{\Delta}\zeta_A^\perp(E_F),\cr\cr
t_-(q^2)&=&-\left(1+\frac{m_A^2}{M^2}\right)
 \frac{E_F}{\Delta}\zeta_A^\perp(E_F),\cr\cr
t_0(q^2)&=&\left(1+\frac{3m_A^2}{M^2}\right)\frac{m_A}{E_F}
\zeta_A^\parallel(E_F)-\frac{2m_A^2}{M^2}\zeta_A^\perp(E_F).  
\end{eqnarray}
On the basis of Eqs.~(\ref{eq:ffav}) the following symmetry relations
between the form factors can be derived
\begin{eqnarray}
  \label{eq:ffrav1}
  &&\frac{M}{M+m_A}\frac{\Delta}{E_F}t_A(q^2)=
\frac{M+m_A}{2E_F}t_{V_1}(q^2)=\left(1+\frac{m_A^2}{M^2}\right)
\frac{\Delta}{E_F}t_+(q^2)
\cr\cr&&\qquad\qquad\qquad\qquad\ \, =
-\left(1-\frac{m_A^2}{M^2}\right)\frac{\Delta}{E_F}t_-(q^2)
=\zeta_A^\perp(E_F),\\\cr
\label{eq:ffrav2}
&&\frac12\left(1-\frac{3m_A^2}{M^2}\right)t_{V_3}(q^2)+
\left(1+\frac{m_A^2}{M^2}\right)\frac{M+m_A}{2E_F}t_{V_1}(q^2)
 \cr\cr&&\qquad\qquad\qquad\qquad\ \,
= 
\left(1-\frac{3m_A^2}{M^2}\right)t_0(q^2)+\frac{2m_A^2}{M^2}t_+(q^2)=
\frac{m_A}{E_F}\zeta_A^\parallel(E_F),\\\cr
&&\qquad\qquad\quad \
t_{V_2}(q^2)=-\frac{m_A^2}{M^2}\left[t_{V_3}(q^2)-2t_{V_1}(q^2)
\right].\nonumber 
\end{eqnarray}

Of particular interest are the 
ratios of form factors which depend only on meson masses and recoil
momentum
following from Eqs.~(\ref{eq:ffrav1}), (\ref{eq:ffrav2}) 
\begin{eqnarray}
  \label{eq:affr}
  \frac{t_A(q^2)}{t_{V_1}(q^2)}&=&\frac{(M+m_A)^2}{2M\Delta},\cr\cr
\frac{t_+(q^2)}{t_+(q^2)+[q^2/(M^2-m_A^2)]t_-(q^2)}
&=&\frac{M^2-m_A^2}{2M\Delta}.
\end{eqnarray}
These relations correspond to the  exact vanishing of helicity amplitudes
$H_+(q^2)$ in the heavy quark and large recoil limit. 
This is similar to
the case of 
$B$ decays to vector light mesons. With the account of equations
(\ref{eq:affr}), the ratio of helicity $\lambda=-1$
and $\lambda=0$ amplitude takes the form
 \begin{equation}
  \label{eq:hpoa}
  \frac{|H_-(q^2)|}{|H_0(q^2)|}=\frac{2\sqrt{q^2}}{M-E_F+\Delta}
\frac{E_F}E\frac{|\zeta_A^\perp(E_F)|}{|\zeta_A^\parallel(E_F)|}.
\end{equation}

\subsection{$\bbox{B}$ decays to  tensor light  mesons}
\label{sec:ts}

The matrix elements of weak current for $B$ decays to tensor mesons
can be decomposed in seven Lorentz-invariant structures
\begin{eqnarray}
  \label{eq:tff1}
  \langle T(p_F)|\bar q \gamma^\mu b|B(p_B)\rangle&=&
\frac{2ig_V(q^2)}{M+m_T} \epsilon^{\mu\nu\rho\sigma}\epsilon^*_{\nu\alpha}
\frac{p_B^\alpha}M  p_{B\rho} p_{F\sigma},\\\cr
\label{eq:tff2}
\langle T(p_F)|\bar q \gamma^\mu\gamma_5 b|B(p_B)\rangle&=&
(M+m_T)g_{A_1}(q^2)\epsilon^{*\mu\alpha}\frac{p_{B\alpha}}M\cr\cr
&&  +[g_{A_2}(q^2)p_B^\mu+g_{A_3}(q^2)p_F^\mu]\epsilon^*_{\alpha\beta}
\frac{p_B^\alpha p_B^\beta}{M^2} \\\cr
\label{eq:tff3}
\langle T(p_F)|\bar q i\sigma^{\mu\nu}q_\nu b|B(p_B)\rangle&=&
2ig_+(q^2)\epsilon^{\mu\nu\rho\sigma}\epsilon^*_{\nu\alpha}
\frac{p_{B\alpha}}M  p_{B\rho} p_{F\sigma},\\\cr
\label{eq:tff4}
\langle T(p_F)|\bar q i\sigma^{\mu\nu}\gamma_5q_\nu b|B(p_B)\rangle&=&
g_{+}(q^2)\left[\epsilon^*_{\alpha\beta}
\frac{p_B^\alpha p_B^\beta}{M} (p_B+p_F)^\mu -
\epsilon^{*\mu\alpha}\frac{p_{B\alpha}}M (M^2-m_T^2)  \right] \cr\cr
&& +g_{-}(q^2)\left[\epsilon^*_{\alpha\beta}
\frac{p_B^\alpha p_B^\beta}{M} q^\mu
-\epsilon^{*\mu\alpha}\frac{p_{B\alpha}}M q^2\right] \cr\cr
& &+g_0(q^2)\epsilon^*_{\alpha\beta}
\frac{p_B^\alpha p_B^\beta}{M^3}\left[(M^2-m_T^2)q^\mu
-q^2 (p_B+p_F)^\mu\right], 
\end{eqnarray}
where $m_T$ and $\epsilon^{\mu\nu}$ are the mass and polarization tensor of 
the tensor meson.

Calculating  the trace in Eq.~(\ref{eq:mxelt}) with 
the  account of (\ref{eq:axiv}) and (\ref{eq:ppf}), (\ref{eq:pol}) in
the heavy quark and large recoil limit, we get  
\begin{eqnarray}
  \label{eq:srt1}
 \langle T(p_F)|\bar q \gamma^\mu b|B(p_B)\rangle&=&2iE_F
 \zeta_T^\perp(E_F)\epsilon^{\mu\nu\rho\sigma}\epsilon^*_{\nu\alpha}
v^\alpha  v_\rho n_\sigma,\\\cr
 \label{eq:srt2}
\langle T(p_F)|\bar q \gamma^\mu\gamma_5 b|B(p_B)\rangle&=&2E_F\Biggl\{
  \zeta_T^\perp(E_F)\left[\epsilon^{*\mu\alpha}v_\alpha-
\epsilon^*_{\alpha\beta} v^\alpha v^\beta \left(\frac{E_F}{\Delta}n^\mu
-\frac{m_T^2}{2E\Delta}v^\mu\right)\right]\cr\cr
&&
+\frac{E}{\Delta}
  \frac{m_T}{E_F}\zeta_T^\parallel(E_F)\epsilon^*_{\alpha\beta} v^\alpha
v^\beta  n^\mu\Biggr\},\\\cr
  \label{eq:srt3}
\langle T(p_F)|\bar q i\sigma^{\mu\nu}q_\nu b|B(p_B)\rangle&=&2iE_F M
\zeta_T^\perp(E_F) \left(1-\frac{m_T^2}{2EM}\right)
\epsilon^{\mu\nu\rho\sigma} \epsilon^*_{\nu\alpha}
v^\alpha n_\rho v_\sigma,\\\cr
\label{eq:srt4}
\langle T(p_F)|\bar q i\sigma^{\mu\nu}\gamma_5q_\nu
b|B(p_B)\rangle&=&2E_F
\Biggl\{M\zeta_T^\perp(E_F)
  \left(1-\frac{m_T^2}{2EM}\right)\Biggl[\epsilon^{*\mu\alpha}v_\alpha-
\epsilon^*_{\alpha\beta} v^\alpha v^\beta \Biggl(\frac{E_F}{\Delta}n^\mu
\cr\cr
&&-\frac{m_T^2}{2E\Delta}v^\mu\Biggr)\Biggr] 
 +\frac{E}{\Delta}\frac{m_T}{E_F}\zeta_T^\parallel(E_F)
  \epsilon^*_{\alpha\beta} v^\alpha
v^\beta \Biggl[(M-E_F)n^\mu\cr\cr
&& -M
    \left(1-\frac{m_T^2}{2EM}\right)v^\mu\Biggr]\Biggr\}.  
\end{eqnarray}
As in the case of vector and axial vector final mesons 
all form factors for $B$ decays to tensor mesons can be expressed 
in terms of two invariant functions
\begin{eqnarray}
  \label{eq:ffte}
  g_V(q^2)&=&\frac{M+m_T}M\frac{E_F}{\Delta}\zeta_T^\perp(E_F),\cr\cr
g_{A_1}(q^2)&=&\frac{2E_F}{M+m_T}\zeta_T^\perp(E_F),\cr\cr
g_{A_2}(q^2)&=&\frac{E_Fm_T^2}{\Delta^2M}\left[2\zeta_T^\perp(E_F)
-\frac{m_T}{E_F}\zeta_T^\parallel(E_F)\right],\cr\cr
g_{A_3}(q^2)&=&2\frac{E_F^2}{\Delta^2}
\left[\frac{Em_T}{E_F^2}\zeta_T^\parallel(E_F)
-\zeta_T^\perp(E_F)\right],\cr\cr
g_+(q^2)&=&-\left(1-\frac{m_T^2}{2EM}\right)\frac{E_F}{\Delta}
\zeta_T^\perp(E_F),\cr\cr
g_-(q^2)&=&\left(1+\frac{m_T^2}{2EM}\right)\frac{E_F}{\Delta}
 \zeta_T^\perp(E_F),\cr\cr
g_0(q^2)&=&\frac{E_FE}{\Delta^2}\left[\frac{m_T^2}{2E^2}\zeta_T^\perp(E_F)-
\frac{m_T}{E_F}\zeta_T^\parallel(E_F)\right].
\end{eqnarray} 
Performing the $m_T/E$ expansion of Eqs.~(\ref{eq:ffte}),
we obtain up to terms of order $m_T^2/M^2$ 
\begin{eqnarray}
  \label{eq:fft}
  g_V(q^2)&=&\frac{M+m_T}M\frac{E_F}{\Delta}\zeta_T^\perp(E_F),\cr\cr
g_{A_1}(q^2)&=&\frac{2E_F}{M+m_T}\zeta_T^\perp(E_F),\cr\cr
g_{A_2}(q^2)&=&\frac{2m_T^2}{M^2}\left(2\zeta_T^\perp(E_F)
-\frac{m_T}{E_F}\zeta_T^\parallel(E_F)\right),\cr\cr
g_{A_3}(q^2)&=&2\left(1+\frac{4m_T^2}{M^2}\right)
\left[\frac{m_T}{E_F}\left(1-\frac{m_T^2}{M^2}\right)\zeta_T^\parallel(E_F)
-\zeta_T^\perp(E_F)\right],\cr\cr
g_+(q^2)&=&-\left(1-\frac{m_T^2}{M^2}\right)
\frac{E_F}{\Delta}\zeta_T^\perp(E_F),\cr\cr
g_-(q^2)&=&\left(1+\frac{m_T^2}{M^2}\right)
 \frac{E_F}{\Delta}\zeta_T^\perp(E_F),\cr\cr
g_0(q^2)&=&-\left(1+\frac{3m_T^2}{M^2}\right)\frac{m_T}{E_F}
\zeta_T^\parallel(E_F)+\frac{2m_T^2}{M^2}\zeta_T^\perp(E_F).  
\end{eqnarray}
As a result the following form factor relations arise from
Eqs.~(\ref{eq:fft})  
\begin{eqnarray}
  \label{eq:ffrt1}
  &&\frac{M}{M+m_T}\frac{\Delta}{E_F}g_V(q^2)=
\frac{M+m_T}{2E_F}g_{A_1}(q^2)=-\left(1+\frac{m_T^2}{M^2}\right)
\frac{\Delta}{E_F}g_+(q^2)
\cr\cr&&\qquad\qquad\qquad\qquad\ \, =
\left(1-\frac{m_T^2}{M^2}\right)\frac{\Delta}{E_F}g_-(q^2)
=\zeta_T^\perp(E_F),\\\cr
\label{eq:ffrt2}
&&\frac12\left(1-\frac{3m_T^2}{M^2}\right)g_{A_3}(q^2)+
\left(1+\frac{m_T^2}{M^2}\right)\frac{M+m_T}{2E_F}g_{A_1}(q^2) 
\cr\cr&&\qquad\qquad\qquad\qquad\ \,
= -
\left(1-\frac{3m_T^2}{M^2}\right)g_0(q^2)-\frac{2m_T^2}{M^2}g_+(q^2)=
\frac{m_T}{E_F}\zeta_T^\parallel(E_F),\\\cr
&&\qquad\qquad\quad 
 g_{A_2}(q^2)=-\frac{m_T^2}{M^2}\left[g_{A_3}(q^2)
-2g_{A_1}(q^2)\right].\nonumber
\end{eqnarray}

Two particular ratios of form factors 
\begin{eqnarray}
  \label{eq:tffr}
  \frac{g_V(q^2)}{g_{A_1}(q^2)}&=&\frac{(M+m_A)^2}{2M\Delta},\cr\cr
\frac{g_+(q^2)}{g_+(q^2)+[q^2/(M^2-m_A^2)]g_-(q^2)}
&=&\frac{M^2-m_A^2}{2M\Delta}
\end{eqnarray}
again lead to the exact vanishing of the corresponding 
helicity amplitude $H_+(q^2)$ in the heavy quark and large recoil limit.
Similarly the ratio of helicity $\lambda=-1$
and $\lambda=0$ amplitudes is given by
 \begin{equation}
  \label{eq:hpot}
  \frac{|H_-(q^2)|}{|H_0(q^2)|}=\frac{2\sqrt{q^2}}{M-E_F+\Delta}
\frac{E_F}E\frac{|\zeta_T^\perp(E_F)|}{|\zeta_T^\parallel(E_F)|}.
\end{equation}

\section{Symmetry relations in the relativistic quark model}
\label{sec:srrm}

We can test the fulfilment of the symmetry relations for the form
factors, arising at large recoil of the final meson, in the
framework of the relativistic quark model based on the quasipotential
approach in quantum field theory. In Refs.~\cite{fgm} and \cite{fg} we
considered exclusive semileptonic and rare radiative
$B$ decays to ground state light mesons. The analysis of these decays
was  performed
near the point of the maximum recoil of the final light meson employing
the expansions both in inverse powers of the heavy $b$-quark mass in
the initial state and in the inverse large recoil momentum of the final 
pseudoscalar or vector meson. The resulting expressions are valid up
to the second order terms in these expansions and for $q^2=0$. The general
formulas (A1), (A4), (A7), (A10) from Ref.~\cite{fgm} and (24) from
Ref.~\cite{fg}  can be applied for decays to ground state light mesons
as  well as for their radial excitations. 
Keeping only the leading terms in $\Lambda_{\rm QCD}/m_b$,
$\Lambda_{\rm QCD}/E_F$ and terms quadratic in
$m/E_F$ we compare the resulting expressions with symmetry relations
(\ref{eq:ffrp}), (\ref{eq:ffrv1}), (\ref{eq:ffrv2}). It is easy to
check that the symmetry relations between the form factors  are
satisfied in our relativistic quark model. As a result one can
determine corresponding invariant functions. In this way we get:

({\it i}) for $B$ decays to pseudoscalar light mesons \nopagebreak
\begin{equation}
  \label{eq:mfp}
  \zeta_P(E_F)=\sqrt{\frac{M}{2E_F}}\int \frac{d^3 p}{(2\pi)^3}
  \bar\psi_P\!\left( {\bf p}+\frac{2\epsilon_q}{E_F+m_P}{\bf\Delta}
  \right) \psi_B({\bf p}),  
\end{equation}

({\it ii}) for $B$ decays to vector light mesons \nopagebreak
\begin{eqnarray}
  \label{eq:mfv}
  \zeta_V^\perp(E_F)&=&\sqrt{\frac{M}{2E_F}}\int \frac{d^3 p}{(2\pi)^3}
  \bar\psi_V\!\left( {\bf p}+\frac{2\epsilon_q}{E_F+m_V}{\bf\Delta}
  \right) \psi_B({\bf p}),\\\cr
\label{eq:mfv1}
\zeta_V^\parallel(E_F)&=&\frac{E_F}E\zeta_V^\perp(E_F).
\end{eqnarray}
Since Eq.~(\ref{eq:mfv1}) coincides with Eq.~(\ref{eq:ffrs1}),
 we can rewrite the relations (\ref{eq:soares2}) in the form:
\begin{eqnarray}
  \label{eq:srr}
  T_1(q^2)&=&A_0(q^2),\cr\cr
{m_V(M-m_V)T_2(q^2)}&=&{(ME_F-m_V^2)A_1(q^2)
-\frac{2M^2\Delta^2}{(M+m_V)^2}A_2(q^2)},\cr\cr
(ME_F+m_V^2)T_2(q^2)
&-&\frac{2M^2\Delta^2}{(M^2-m_V^2)}T_3(q^2)={m_V(M+m_V)A_1(q^2)},
\end{eqnarray}
thus recovering the form factor relations of Ref.~\cite{soares2} near
the maximum recoil of the final vector meson.  

In Ref.~\cite{efgt} we considered rare radiative $B$ decays to
orbitally excited $K$ mesons up to the second order terms of heavy quark and
large recoil momentum expansions. We further retain only the
contributions  of leading
order in  $\Lambda_{\rm QCD}/m_b$, $\Lambda_{\rm QCD}/E_F$ and of
second order in $m/E_F$ in Eqs.~(29), (33)
and (37) of Ref.~\cite{efgt}. Then we perform a  similar
analysis for the semileptonic $B$ decays to orbitally excited scalar,
axial vector and tensor light mesons. The symmetry relations
(\ref{eq:ffrs}), (\ref{eq:ffrav1}), (\ref{eq:ffrav2}),
(\ref{eq:ffrt1}), (\ref{eq:ffrt2}) are valid in our model and
invariant functions can be obtained. Thus we find:

({\it iii}) for $B$ decays to scalar light mesons \nopagebreak
\begin{eqnarray}
\label{eq:mfs}
\zeta_S(E_F)&=&\frac{\Delta}{E_F+m_S}\frac13
\sqrt{\frac{M}{2E_F}}\int \frac{d^3 p}{(2\pi)^3}
  \bar\psi_S\!\left( {\bf p}+\frac{2\epsilon_q}{E_F+m_S}{\bf\Delta}
  \right)\cr\cr
&&\times\left[-3(E_F+m_S)\frac{{\bf p}\cdot {\bf\Delta}}{p\Delta^2}
  -\frac{2p}{\epsilon_q(p)+m_q}\right]\psi_B({\bf p}),
\end{eqnarray}

({\it iv}) for $B$ decays to axial vector light meson with $j=1/2$ \nopagebreak
\begin{eqnarray}
  \label{eq:mfa1}
  \zeta^\perp_{A(1/2)}(E_F)&=&\frac{\Delta}{E_F+m_{A(1/2)}}\frac13
\sqrt{\frac{M}{2E_F}}\int \frac{d^3 p}{(2\pi)^3}
  \bar\psi_{A(1/2)}\!\left( {\bf p}
+\frac{2\epsilon_q}{E_F+m_{A(1/2)}}{\bf\Delta}
  \right)\cr\cr
&&\times\left[-3(E_F+m_{A(1/2)})\frac{{\bf p}\cdot {\bf\Delta}}{p\Delta^2}
  -\frac{2p}{\epsilon_q(p)+m_q}\right]\psi_B({\bf p}),\\\cr
\label{eq:mfa12}
\zeta_{A(1/2)}^\parallel(E_F)&=&-\frac{E_F}E\zeta_{A(1/2)}^\perp(E_F),
\end{eqnarray}

({\it v}) for $B$ decays to axial vector light mesons with $j=3/2$ \nopagebreak
\begin{eqnarray}
  \label{eq:mfa2}
  \zeta^\perp_{A(3/2)}(E_F)&=&\frac{\Delta}{E_F+m_{A(3/2)}}\frac1{3\sqrt{2}}
\sqrt{\frac{M}{2E_F}}\int \frac{d^3 p}{(2\pi)^3}
  \bar\psi_{A(3/2)}\!\left( {\bf p}
+\frac{2\epsilon_q}{E_F+m_{A(3/2)}}{\bf\Delta}
  \right)\cr\cr
&&\times\left[-3(E_F+m_{A(3/2)})\frac{{\bf p}\cdot {\bf\Delta}}{p\Delta^2}
  +\frac{p}{\epsilon_q(p)+m_q}\right]\psi_B({\bf p}),\\\cr
\label{eq:mfa22}
\zeta_{A(3/2)}^\parallel(E_F)&=&2\frac{E_F}E\zeta_{A(3/2)}^\perp(E_F),
\end{eqnarray}

({\it vi}) for $B$ decays to tensor light mesons \nopagebreak
\begin{eqnarray}
  \label{eq:mft}
  \zeta^\perp_{T}(E_F)&=&\frac{m_T}{E_F+m_T}\frac1{\sqrt{3}}
\sqrt{\frac{M}{2E_F}}\int \frac{d^3 p}{(2\pi)^3}
  \bar\psi_{T}\!\left( {\bf p}
+\frac{2\epsilon_q}{E_F+m_{T}}{\bf\Delta}
  \right)\cr\cr
&&\times\left[-3(E_F+m_{T})\frac{{\bf p}\cdot {\bf\Delta}}{p\Delta^2}
  +\frac{p}{\epsilon_q(p)+m_q}\right]\psi_B({\bf p}),\\\cr
\label{eq:mft2}
\zeta_{T}^\parallel(E_F)&=&\frac{E_F}E\zeta_{T}^\perp(E_F).
\end{eqnarray}
Here $\psi_B$ and $\psi_F$ are the $B$ meson and light meson
wave functions.
We used the HQET $js$-coupling scheme for parametrizing axial vector
state wave functions, which is convenient for the description of axial
vector orbital excitations of $K$ mesons \cite{efgt}. 
Instead, for axial vector excitations
of $\pi,\rho$ mesons the $LS$-coupling scheme should be used  where
${}^1P_1$ and ${}^3P_1$ states are
the linear combinations of $A(1/2)$ and $A(3/2)$ states.

The ratio of the helicity $\lambda=-1$ and $\lambda=0$  amplitudes 
(\ref{eq:hpmr}), (\ref{eq:hpoa}), (\ref{eq:hpot}) with
the account of quark model relations (\ref{eq:mfv1}),
(\ref{eq:mfa12}), (\ref{eq:mfa22}), (\ref{eq:mft2}) for small $q^2$  will 
depend   on meson masses and energies only
\begin{equation}
  \label{eq:hphmr}
  \frac{|H_-(q^2)|}{|H_0(q^2)|}=\frac{2\sqrt{M^2+m_V^2-2ME_F}}{M-E_F+\Delta}
\left\{
\begin{array}{c}
1\cr\cr
1\cr\cr
1/2\cr\cr
1
\end{array}
\hskip 1.5cm
\begin{array}{l}
F=V \cr\cr F=A(1/2)\cr\cr F=A(3/2)\cr\cr  F=T
\end{array}\right. .
\end{equation}
The ratio of the helicity $\lambda=+1$ and $\lambda=0$ amplitudes
vanishes,
$|H_+(q^2)|/|H_0(q^2)|=0$, for all above decays in the heavy quark and large
recoil limit. 
These predictions can be tested in charmless nonleptonic $B$ decays to
two  light final mesons in the factorization approximation.

In Tables~\ref{tab:1} and \ref{tab:2} we give our predictions for the
values  of invariant
functions $\zeta_{P,S}$, $\zeta_{V,A,T}^\perp$ and
$\zeta_{V,A,T}^\parallel$ parametrizing $B$ decays to light meson
states at the maximum recoil  of the
final meson where $E_F^{\rm max}=(M^2+m^2)/(2M)$ and $q^2=0$. These
values can be used for evaluating the decay rates.

Recently, in Refs.~\cite{bfs,bb} the exclusive $B\to K^*\gamma$ decay
branching fractions were calculated with the complete account of
next-to-leading order QCD effects and to leading order in the heavy
quark and large recoil limit. A sizable enhancement of the
coefficient ${\cal C}_7$  at next-to-leading order was observed. 
Ref.~\cite{bfs} gives
\begin{equation}
  \label{eq:cnlo}
  |{\cal C}_7|^2_{\rm NLO}/|{\cal C}_7|^2_{\rm LO}\approx 1.78,
\end{equation}
and the close value $1.6$ was found in Ref.~\cite{bb}. The
decay branching fraction in the above limit should be proportional to
the square of the transverse invariant function
$\zeta_{K^*}^\perp(E_F^{\rm max})$.  In
order to evaluate $\zeta_{K^*}^\perp(E_F^{\rm max})$ the predictions
of QCD sum  rules \cite{balbr}
for the decay form factors $T_1^{K^*}(0)$ and $V^{K^*}(0)$ were
used. The estimate of Ref.~\cite{bfs} yields 
$\zeta_{K^*}^\perp(E_F^{\rm max})=0.35\pm0.07$,
which leads to the branching fraction 
\begin{equation}
  \label{eq:brbkgt}
  BR(B\to K^*\gamma)=(7.9^{+1.8}_{-1.6})\times 10^{-5}
\left(\frac{\tau_B}{1.6\ ps}\right)\left(\frac{m_b}{4.6\ {\rm
      GeV}}\right)^2\left(
\frac{\zeta_{K^*}^\perp(E_F^{\rm max})}{0.35}\right)^2. 
\end{equation}
This value is considered in \cite{bfs}  to be nearly twice as large as
the current experimental data 
\begin{equation}
  \label{eq:brbkge}
  BR(B^0\to K^{*0}\gamma)=\left\{\begin{array}{c}
(4.55\pm 0.70\pm 0.34)\times 10^{-5}\cr
(4.96\pm 0.67\pm 0.45)\times 10^{-5}\cr
(4.39\pm 0.41\pm 0.27)\times 10^{-5}
\end{array}
\hskip 1.5cm
\begin{array}{r}
{\rm CLEO \cite{cleo}}\cr
{\rm Belle \cite{belle}}\cr
{\rm BABAR \cite{babar}}
\end{array}\right.\end{equation}
\begin{equation}
\label{eq:brbkgep}\hskip 0.3cm
 BR(B^+\to K^{*+}\gamma)=\left\{\begin{array}{c}
(3.76\pm 0.86\pm 0.28)\times 10^{-5}\cr
(3.89\pm 0.93\pm 0.41)\times 10^{-5}
\end{array}
\hskip 1.8cm
\begin{array}{r}
{\rm CLEO \cite{cleo}}\cr
{\rm Belle \cite{belle}}
\end{array}\right. .
\end{equation}
In Ref.~\cite{bb} the value
$F_{K^*}(0)=\zeta_{K^*}^\perp(E_F^{\rm max})=0.38\pm0.06$ is used and
the estimates
\begin{equation}
  \label{eq:brbb}
  BR(B^-\to K^{*-}\gamma)=7.45\times 10^{-5}, \qquad
BR(B^0\to K^{*0}\gamma)=7.09\times 10^{-5}
\end{equation}
are obtained
for central input parameters. The authors of \cite{bb} claim, that
these results are compatible with experimental measurements taking
the sizable uncertainties into account, even though the central
theoretical values appear to be somewhat high.

However,
 $\zeta_{K^*}^\perp(E_F^{\rm max})$ 
determined from QCD
sum rules contains $\Lambda_{\rm QCD}/m_b$ and $\Lambda_{\rm QCD}/E_F$
corrections which were neglected in deriving (\ref{eq:cnlo}). As our
model estimates show,
 $1/m_b$ corrections can give contributions of
$10-15\%$ to the form factors. Indeed,  with the account of
such corrections we have
  $T_1^{K^*}(0)=0.32(3)$ in our model \cite{fg}
while in the large recoil limit 
$T_1^{K^*}(0)\to\zeta_{K^*}^\perp(E_F^{\rm max})=0.28$ (see
Table~\ref{tab:1}).  If we substitute the latter value in
Eq.~(\ref{eq:brbkgt}), we get
\begin{equation}
  \label{eq:bkgn}
  BR(B\to K^*\gamma)=(5.0^{+1.8}_{-1.7})\times 10^{-5},
 \end{equation}
while the 
branching fractions (\ref{eq:brbb}) would be given  
in our model by
\begin{equation}
  \label{eq:bkgbb}
  BR(B^-\to K^{*-}\gamma)=(4.1^{+1.5}_{-1.2})\times 10^{-5},
\qquad BR(B^0\to K^{*0}\gamma)=(3.9^{+1.5}_{-1.2})\times 10^{-5}
\end{equation}
in agreement with experimental data (\ref{eq:brbkge}) and (\ref{eq:brbkgep}).

The similar substitution of our prediction for
$\zeta_\rho^\perp(E_F^{\rm max})=0.24$ (see Table~\ref{tab:1}) 
instead of the QCD sum
rule \cite{balbr} value $T_1^\rho(0)=0.29\pm 0.04$ in the expression
for $BR(B\to \rho\gamma)$ with the complete account of
next-to-leading order QCD corrections in the heavy quark and large
recoil limit \cite{bb} (see also \cite{ap}) yields
\begin{equation}
  \label{eq:brbtr}
  BR(B^-\to\rho^-\gamma)=(1.1^{+0.5}_{-0.4})\times 10^{-6}.
\end{equation}

\section{Conclusions}
\label{sec:conc}

In this paper we derived the form factor relations for $B$ decays to
light mesons arising in the heavy quark and large recoil energy
limits. The decays both to ground state and radially
and orbitally excited light mesons were considered. The main attention
was devoted to the complete accounting for corrections of second order
in the ratio of the light meson mass to the large recoil
energy.  Such corrections are especially
important for decays to excited light mesons, since their masses are
of order of the charmed quark mass. The correction to the effective
Lagrangian  quadratic in the final meson mass has been obtained. It
was found that this correction does not violate the symmetry
of the leading order Lagrangian, since it has the same Dirac structure
as the leading contribution. Therefore the
inclusion of  corrections, which are
quadratic in the ratio of the final meson mass to the large recoil
energy,  to the weak current matrix elements does
not lead to the introduction of additional invariant functions. Their
inclusion requires a more accurate consideration of the decay
kinematics, keeping all final meson mass contributions. The heavy
quark and large recoil symmetries substantially constrain the number
of independent form factors. Thus,
in this limit for $B$ decays to pseudoscalar 
(scalar) light mesons three decay form factors can be parametrized by one
invariant function $\zeta_{P,S}(q^2)$, and for $B$ decays to vector
(axial-vector, tensor) light mesons seven decay form factors can be
expressed through two invariant functions $\zeta_{V,A,T}^\perp(q^2)$
and $\zeta_{V,A,T}^\parallel (q^2)$ for each decay, respectively. This
establishes relations between decay form factors at the large recoil of the
final light meson, which are
obtained with the complete account of second order corrections in the
light to $B$ meson mass ratio.     

The important consequence of these equations are the well known
Isgur-Wise \cite{iw} relations (\ref{eq:iwr}) between form factors of
semileptonic and rare radiative $B$ decays, which were originally
obtained for small values of the recoil momentum, and now they are
established near the point of maximum recoil if all contributions
quadratic in final meson mass are included. The 
relations (\ref{eq:soar}) obtained by Soares 
in the constituent quark model \cite{soares}
also follow
from the large recoil symmetry, while the fulfilment of the other ones
(\ref{eq:soares2}) requires additional relation between the transverse and
longitudinal functions (\ref{eq:ffrs1}).

We tested the fulfilment of the large recoil symmetry relations
between the $B$ decay form factors
in the framework of the relativistic quark model \cite{fg,fgm,efgt}
based on the quasipotential approach in quantum field theory. It was
found that they are exactly satisfied in the appropriate limits
and corresponding invariant functions were determined. 
The additional relation (\ref{eq:ffrs1}) is also satisfied in our model.
Estimates of the heavy-to-light form factors at $q^2=0$ in our model
show that their values with the account of $1/m_b$ corrections can
differ from the ones in the heavy quark and large recoil
limit by $10-15\%$. E.~g. the form factor $T_1^{K^*}(0)$
responsible for the rare radiative $B\to K^*\gamma$ decay is reduced
by  $\sim 12\%$  in the large recoil limit. Such reduction of the form
factor almost compensates the enhancement of this decay rate by
next-to-leading order QCD corrections calculated in the same limit
\cite{bfs,bb}.

\acknowledgements
The authors express their gratitude to A. Ali, 
M. M\"uller-Preussker, V. Savrin and H. Toki 
for support and discussions.
D.E. acknowledges  the support provided
to him by the Ministry of Education and Science and Technology 
of Japan  (Monkasho) for his
work at RCNP of Osaka University. 
He is also grateful to H. Toki for the kind hospitality at RCNP.
Two of us (R.N.F and V.O.G.)
were supported in part by the {\it Deutsche
Forschungsgemeinschaft} under contract Eb 139/2-1, {\it
 Russian Foundation for Fundamental Research} under Grant No.\
00-02-17768 and {\it Russian Ministry of Education} under Grant
No. E00-3.3-45.

\begin{table}[htbp]
  \caption{Values of invariant functions $\zeta_P$, $\zeta_V^\perp$ 
and $\zeta_V^\parallel$ parametrizing $B$ decays to ground state and radially
excited pseudoscalar and vector light mesons at the maximum recoil of the 
final light meson calculated in the relativistic quark model.     } 
\begin{center}
    \begin{tabular}{lc|lcc}
Decay& $\zeta_P(E_F^{\rm max})$&Decay&$\zeta_V^\perp(E_F^{\rm max})$ 
&$\zeta_V^\parallel(E_F^{\rm max})$\cr
\hline
$B\to \pi$&0.19&$B\to\rho$ &0.24&0.25\cr
$B\to K$&  0.26&$B\to K^*$ &0.28&0.29\cr
$B\to\pi(1430)$&0.24&$B\to\rho(1450)$&0.25&0.27\cr
$B\to K(1460)$&0.19&$B\to K^*(1410)$&0.20&0.21 
    \end{tabular}
   
    \label{tab:1}
  \end{center}
\end{table}
\begin{table}[htbp]
  \caption{Values of invariant functions $\zeta_S$, $\zeta_{A,T}^\perp$ 
and $\zeta_{A,T}^\parallel$ parametrizing $B$ decays to orbitally
excited ($P$-wave) scalar, axial vector and tensor light mesons at the maximum 
recoil of the final light meson calculated in the relativistic quark model.} 
\begin{center}
    \begin{tabular}{lc|lcr|lcc}
Decay& $\zeta_S(E_F^{\rm max})$&Decay&$\zeta_A^\perp(E_F^{\rm max})$
&$\zeta_A^\parallel(E_F^{\rm max})$
&Decay&$\zeta_T^\perp(E_F^{\rm max})$&$\zeta_T^\parallel(E_F^{\rm max})$\cr
\hline
$B\to a_0(1450)$&0.14&$B\to b_1(1235)$ &0.08&$-0.08$&&&\cr
$B\to K_0^*(1430)$&0.14&$B\to K^*_1(1270)$ &0.14&$-0.15$
& &&\cr
&&$B\to a_1(1260)$ &0.23&0.49&$B\to a_2(1320)$ &0.25&0.27 \cr
&&$B\to K_1(1400)$ &0.20&0.43&$B\to K^*_2(1430)$ &0.27&0.29\cr
    \end{tabular}
   
    \label{tab:2}
  \end{center}
\end{table}

\end{document}